\documentstyle[12pt,aas2pp4,psfig]{article}

\newcommand{\be}{\begin{equation}}
\newcommand{\ee}{\end{equation}}
\newcommand{\bea}{\begin{eqnarray}}
\newcommand{\eea}{\end{eqnarray}}
\begin{document}


\newcommand{\VV}[1]{\mbox{\boldmath{$#1$}}}   
\newcommand{\dif}[2]{\frac{\partial #1}
                          {\partial #2}}      %
\newcommand{\dift}[2]{\partial #1 /
		      \partial #2}           
\newcommand{\diff}[2]{\frac{\partial^{2} #1}%
			   {\partial #2^{2}}} 
\newcommand{\Dif}[2]{\frac{\mathrm{d} \/#1}
                    {\mathrm{d} \/#2}}        %
\newcommand{\Dift}[2]{\mathrm{d} \/#1 /
		     \mathrm{d} \/#2}         
\newcommand{\nab}[0]{\vec{\nabla}}            
\newcommand{\Div}[1]{\vec{\nabla}\cdot\/#1}   
\newcommand{\Dim}[1]{\mbox{$\,\mathrm{#1}$}}  
\newcommand{\eqnref}[1]{(\ref{#1})}           
\newcommand{\fref}[1]{\ref{#1},               %
            Seite \pageref{#1}}               %
\newcommand{\argu}[1]{\! \left( #1 \right)}   
\newcommand{\Achtung}[1]{\emph{#1} 
            \marginpar{\Huge\textbf{!}}}      
\newcommand{\code}[1]{\texttt{#1}}            
\newcommand{\file}[1]{\texttt{#1}}            
\newcommand{\class}[1]{\texttt{#1}}           
\newcommand{\funct}[1]{\texttt{#1}}           
\newcommand{\prog}[1]{\texttt{#1}}            
\newcommand{\ds}[0]{\displaystyle}            
\newcommand{\ts}[0]{\textstyle}               

\newcommand{\invh}[1]
   {\newlength{\hei}%
    \newlength{\dep}%
    \settoheight{\hei}{#1}%
    \settodepth{\dep}{#1}%
    \addtolength{\hei}{\dep}%
    \rule[\dep]{0pt}{\hei}%
   }
\newcommand{\invw}[1]
   {\newlength{\wid}%
    \settowidth{\wid}{#1}%
    \rule{\wid}{0pt}%
   }

\newcommand{\aref}[1]%
	{(\/\textit{Fig.~\ref{#1}}\/)} 	      

\newcommand{\bref}[2]%
	{(\/\textit{Fig.~\ref{#1}(#2)}\/)}    
\newcommand{\ind}[1]{_{\rm{#1}}}              

\title{The critical velocity effect as a cause for the H$\alpha$ emission
from the Magellanic stream}

\author{C.~Konz, H.~Lesch, G.T.~Birk, and H.~Wiechen}
\affil{Centre for Interdisciplinary Plasma Science (CIPS) \\
Institut f{\"u}r Astronomie und Astrophysik der Universit{\"at}
M{\"u}nchen\\ Scheinerstra{\ss}e 1, D-81679 M{\"u}nchen, Germany}
\begin{abstract}
\noindent
Observations show significant H$\alpha$-emissions in the Galactic halo near
the edges of cold gas clouds of the Magellanic Stream. The source for the
ionization of the cold gas is still a widely open question. In our paper we
discuss the critical velocity effect as a possible explanation for the
observed H$\alpha$-emission. The critical velocity effect can yield a fast
ionization of cold gas if this neutral gas passes through a magnetized plasma
under suitable conditions. We show that for parameters that are typical
for the Magellanic Stream the critical velocity effect has to be considered as
a possible ionization source of high relevance.
 \end{abstract}


\section{Introduction}
The Magellanic Stream (MS) consists of a narrow band of high-velocity
neutral hydrogen gas (HI) which stretches over almost 100$^{o}$ on the sky,
trailing the Magellanic Clouds in their orbit around the Milky Way
(Mathewson, Cleary \& Murray 1974; Mirabel, Cohen \& Davies 1979). The Stream was observed
at 21 cm. Thereby a chain of clouds connected by a gas of lower density
was identified.
The clouds have been labeled MS I through MS VI (Mathewson, Schwarz \& Murray 1977). They
are characterized by a high density concentration with a relatively sharp gradient
on the leading edge, which is defined by the direction of the proper motion of
the Large Magellanic Cloud (Jones, Klemola \& Lin 1994). The Stream has no known
stellar component (Mathewson et al. 1979) and its distance to the Milky Way
is unknown, although its leading end (MS I) connects to the Magellanic Clouds
and is presumably located at a distance of 50- 60 kpc. Moore and Davis (1994)
estimate the distance of the tip of the Stream (MS VI) to be 20 kpc, whereas Gardiner,
Sawa \& Fujimoto (1994) give a distance value of 60 kpc.

Bregman (1979) considered the MS as the hydrodynamic
result of the tidal interaction between the Magellanic Clouds and
the gaseous halo of the Milky Way (Murai \& Fujimoto 1980;
Lin \& Lynden-Bell 1982).
This tidal scenario is supported by an extension to the model (Moore \& Davis 1994)
which was able to reproduce the kinematics of the Stream  by including the
stripping associated with an extended ionized disk and drag due to a diffuse halo.
Nevertheless, the models for the origin of the
MS remain controversial (Wakker \& van Woerden 1997; Murali 2000), so that the inferences about halo
gas properties are correspondingly uncertain.

What do we know from observations about the characteristic values for
the density and temperature of the MS?

The northern part
of the MS (MS III - MS MVI) has been surveyed with high sensitivity by Mirabel, Cohen \& Davies (1979)
in order to study the small scale structure. Five regions were investigated, which
are characterized by narrow filaments running parallel to the main direction
of the Stream, and small cloud-like concentrations. These clouds have typical
masses of 15 $D^2 M_\odot$
(here D is their distance in kpc). The authors conclude
that the clouds appear to be unstable over the lifetime of the Stream as a whole
and some external confinement mechanism seems to be necessary to preserve the
observed small scale structure. This could be provided by a hot plasma of density
$\sim 4\cdot 10^{-4}\, {\rm cm}^{-3}$ at a temperature of $10^6\, $K.

Weiner \& Williams (1996) detected faint H$\alpha$-emission from several
points of the MS, i.e. the leading edges of the HI clouds MS II, MS III and MS IV.
The emission is strongest (emission measures EM: 0.5-1 ${\rm cm}^{-6}$ pc)
on the sharp leading-edge density gradient.
They suggest that ram pressure from the halo gas plays an important
 role in stripping the
Stream out of the Magellanic Clouds and they suggest the existence of a
relatively
dense ($n_H\sim 10^{-4}\,  {\rm cm}^{-3}$) and almost fully ionized
gas with a temperature of $\sim 1.7 \times 10^6\, $K
in the Galactic halo at a radius of 50 kpc.

Weiner and Williams (1996) rule out photoionization as a cause for the H$\alpha$-emission.
They argue that the HI column
densities in all observed fields are optically thick to ionizing radiation
which means that any uniform ionizing
flux should produce ionization rates and H$\alpha$ emission which are roughly constant from
field to field, which is not observed.
The low number of required UV-photons ($10^{48}\, {\rm s}^{-1}$) could,
in principle,
be provided by a few OB stars located  $\sim 100$ pc away from the
observed field. However, such stars that
would be easily detectable by the associated luminous HII-region are not
observed.

Murali (2000) recently discussed some theoretical implications.
He reexamined the interaction of the MS with the
surrounding medium at large distances in the halo of the Galaxy. He was able to show
that heating dominates over drag forces. Due to their high relative velocities
($\ge  200$ km ${\rm s}^{-1}$)
the HI-clouds should evaporate if the density
of the halo was too high. Thus, he concluded a limit for the halo density
at 50 kpc to be less than $10^{-5}\, {\rm cm}^{-3}$.

The strength and orientation of the magnetic field in the halo of our
Galaxy is still a matter of debate (e.g. Vallee 1998). According to
Beuermann, Kanbach \& Berkhuijsen (1985) the scale height of the field is
$\sim 6 $kpc. The distance of the MS is several tens of kpc. Given a
disk magnetic field of about $6\mu$G and taking into account local
field compression by the MS clouds
the magnetic field in the MS is expected to be of the order of
several $\mu$G. Amplifications of magnetic fields
by infalling gas clouds is well-known from high velocity clouds
(e.g. Crutcher 1999; Vallee 1998).

The interaction of the high velocity clouds (HVC) of the MS and the medium
of the galactic halo is an example for the interaction of a fully ionized
plasma and a cold weakly ionized gas.
Such a situation has been considered by Alfv\'en (1954)
in connection with his model for the origin of the solar system.
He proposed that when  a neutral gas streams through a plasma across
magnetic field lines at a velocity that exceeds a certain value $v_c$
a discharge like process can occur in which the neutral gas component
is rapidly ionized.
The critical velocity is proportional to the ionization potential
and inversely proportional to the square root of the mass of the neutral atom.
This effect is usually considered in the context of plasmas in which
the ion-neutral collision rate is negligible. This is exactly the case
in the interaction regions of the MS HI-clouds and the galactic halo.
Given the problematic issue of lacking ionization sources for the
H$\alpha$, the critical velocity effect (CVE) presents an
interesting fundamental plasma
process that is capable of providing the required ionization. 
The CVE occurs when a neutral gas streams through a plasma across magnetic 
field lines with a velocity that exceeds a critical value measured by the 
ionization potential of the neutrals. The kinetic energy of the flow is 
partly converted to heating the electrons which can efficiently ionize 
the neutral gas component.

It is the aim of our contribution to discuss the CVE in the
astrophysical system MS-Galactic halo as an alternative
ionization mechanism far away from standard ionization sources like stars.

In Sect. 2 we review the basic conditions for the CVE to work. We give limits
for the magnetic field strength, neutral gas, plasma density, and plasma
temperature. Sect. 3 presents the application to the ionization of the
MS cloudlets, followed by Sect. 4 which contains the discussion and
some conclusions, in particular, on the general applicability of the
CVE-process in astrophysical plasma-neutral gas interactions.

\section{The Critical Velocity Effect}

The interaction of a plasma and a neutral gas in the presence of a
magnetic field can result in the ionization of the neutral component
if the relative velocity component across the magnetic field $v_{rel}$
exceeds a critical value
\be
v_{rel}>v_c=\epsilon^{-1} \left(\frac{2e\Phi}{m_n}\right)^\frac{1}{2}\label{1}
\ee
where $e$, $\Phi$, and $m_n$ denote the elementary charge, the ionization
potential of the neutrals, and the neutral mass.
This is the so-called critical velocity effect (CVE) (Alfv\'en 1954;
Raadu 1978, 1981; Petelski 1981). The efficiency is
parameterized by $\epsilon$. Formisano et al. (1982) have shown on the basis
of quasilinear calculations that the efficiency can drop to $\epsilon\approx
0.025$ if the ionized neutrals, i.e. the newly born ions,
have sufficient time to isotropize in the form of a ring-like velocity
distribution in the plane perpendicular to the magnetic field.
The rather poor efficiency, or equivalently, the necessary
very high relative velocities,
can be avoided for two cases. First, if the ionization
frequency is much higher than the ion plasma frequency
(Formisano et al. 1982), and second,
if the growth rate of the excited lower-hybrid drift instability $\gamma_{lhd}$
that  heats the ionizing electrons is sufficiently high (Brenning 1985),
i.e. $\gamma_{lhd} \gg \Omega_i$
( $\Omega_i=eB/m_ic$ is the ion gyro frequency; $e$, $B$, $m_i$, and
$c$ denote the elementary charge, the magnetic field strength, the ion mass,
and the speed of light, respectively).
The first
alternative is not of interest in the present context and, in fact,
should only
be of importance in dense laboratory plasmas. The latter condition can
be written as (cf. Brenning 1985)
\be
\left(\frac{m_e}{m_n}\right)^\frac{1}{2}\ll \frac{\omega_e}{\Omega_e}
< \frac{c}{v_{rel}}\left(\frac{m_e(1+\beta_e)}{m_i}\right)^\frac{1}{2}
\label{2}
\ee
where $\omega_e=(4\pi n_e e^2/m_e)^{1/2}$
($n_e$ is the electron particle density) and $\Omega_e=eB/m_ec$
denote the electron plasma and gyro frequencies and
$\beta_e = 8\pi n_ekT_e/B^2$
($k$ and $T_e$ are the Boltzmann constant and the electron temperature)
is the electron plasma beta. For high-$\beta$ plasmas
efficient ionization is possible for electron temperatures $kT_e >\frac{1}{2}
m_i v^2_{rel}$.
For low-$\beta$ plasmas the relative
flow velocity $v_{rel}$ has to be sub-Alfv\'enic, $v_{rel}
< v_A=B/(4\pi n_i m_i)^{1/2}$, due to the second inequality.
Here and in the following we assume quasineutrality $n_e=n_i=n$.
In the low-$\beta$ case Eq.(\ref{2}) can be reformulated
\be
7 \ {\rm G}^{-1}{\rm cm}^{-\frac{3}{2}}\ll\frac{n^\frac{1}{2}}{B}<
7 \ \frac{c}{v_{rel}}{\rm G}^{-1}{\rm cm}^{-\frac{3}{2}},\label{3}
\ee
if we choose the ion and neutral masses equal to the proton mass, i.e.
we deal with a partially ionized hydrogen plasma, or
\be
v_A\ll c<\frac{v_A}{v_{rel}}c.\label{4}
\ee
Fig.~1 illustrates the constraints for the
magnetic field strength and the particle density
that guarantee efficient heating for $v_{rel}=10^{-2}c$ and
for $v_{rel}=10^{-4}c$, for example. In the first case,
the region between the solid line and the dotted one represents the
combinations of particle density and magnetic field that allows for the CVE
to operate efficiently. In the latter case ($v_{rel}=10^{-4}c$),  the same
holds for the region between the solid line and
the dashed one.


Up to now we concentrated on the hydrogen component which is dominant
in the interstellar and intergalactic medium.
In the context of the 'missing ionized helium' puzzle (Domg\"orgen 1997;
Reynolds \& Tufte 1995),
it is of interest to discuss the influence
of the CVE on the helium component. In fact, the critical velocity (Eq.~1)
for the onset of the CVE is reduced  by the factor
$\sqrt{(24.6/13.7)/4}\approx 0.67$ as compared to the pure hydrogen case.
However, the efficiency of the ionization is drastically reduced for
two reasons.  First, the resonant heating of the ionizing electrons
depends on the growth rate of the modified two-stream instability which
scales as the ion plasma frequency $\omega_i$. This frequency is much smaller
for the heavier and much rarer helium component than for the hydrogen one.
Consequently, the helium ions are expected to isotropize very fast
and thus, cannot be ionized efficiently, i.e. $\epsilon \approx 0.025$,
(see discussion by Raadu 1978; Formisano et al. 1982; Brenning 1985);
Second, ionization is possible only, if the Townsend criterion
$\nu_{ion}>\tau_{e-i}^{-1}$ if fulfilled where $\nu_{ion}$ and $\tau_{e-i}$ are the
ionization frequency and the time an electron has to ionize an ion.
Thus, the neutral gas density has to exceed a critical value
$n_n>n_c=[\tau_{e-i}\sigma_{ion}(E_e)\sqrt{2E_e/m_e}]^{-1}$ (where
$E_e$ is the average energy of a single ionizing electron and $\sigma_{ion}$
is the ionization cross section). Whereas the ionization cross section
for hydrogen and helium do not differ that much (see Mitchner \& Kruger 1973)
for our applications the particle density $n_n$
of the neutral helium is much lower than the one of HI.

\section{Ionization of the Cloudlets in the Magellanic Stream}

In the context of HVC's in the Galactic halo the CVE should work very
efficiently. Concerning the high-velocity cloudlets in the MS,
e.g.~MS I-VI, several authors have established a set of values for the relevant
plasma parameters as the particle density~\(n\), temperature~\(T\), and magnetic
field strength~\(B\). Doppler measurements of H\( \alpha \)-emission lines show
typical velocities~\(v^{HVC}\) of the cloudlets in the range of \( 110 \Dim{km}
\Dim{s^{-1}} - 220 \Dim{km} \Dim{s^{-1}} \) (Weiner \& Williams 1996; Murali
2000) for the central part of the MS and up to \( 350 \Dim{km}
\Dim{s^{-1}} \) for cloudlets at the tip of the Stream (Mirabel et al. 1979).
The cloudlets show a rather low temperature of \( T \ind{n}^{HVC} \approx 10^{4} \Dim{K}
\) in a hot ambient halo gas at a temperature of \( T \ind{e} ^{halo} \approx 2
\cdot 10^{6} \Dim{K} \) (Weiner \& Williams 1996). The particle density of the
HVC's lies around \( n ^{HVC} \approx 5 \cdot 10^{-2} \Dim{cm^{-3}} \) (Cohen
1982). However, the particle density of the ambient halo gas cannot be
determined so easily. Direct measurements of the halo density are not possible
so far since only column densities can be measured and the problem of the exact
distance of the MS still remains unsolved. The only way to get an
estimate of the halo density consists in the deduction of the density via the
use of hydrodynamic models for the HVC's. From a rather simple drag model Weiner
and Williams deduce a halo density of the order \( n^{halo} \approx 10^{-4}
\Dim{cm^{-3}} \) while Murali finds a maximum value of \( 10^{-5} \Dim{cm^{-3}}
\) from lifetime considerations. The latter value, however, does not seem very
likely considering the fact that the particle density of the intergalactic
medium already shows a minimum value of \( 10^{-6} \Dim{cm^{-3}} \). By assuming
a pressure equilibrium between the cloudlets and the ambient halo gas, Mirabel et
al. find particle densities between \( 7 \cdot 10^{-5} \Dim{cm^{-3}} \) and \( 7
\cdot 10^{-4} \Dim{cm^{-3}} \) depending on the distance of the MS. \\
The above data are shown as single data points in Fig.~1 assuming a magnetic
field strength in the halo of \( B^{halo} \approx 3 \mu \Dim{G} \) (Beuermann et
al. 1985). For the mentioned halo plasma parameters the electron plasma beta~\(
\beta \ind{e} \) is much smaller than unity such that equation~\eqnref{3}
holds. The two dash-dotted lines in Fig.~1 mark the lower boundaries of the
regions of efficient
electron heating for the cases \( v \ind{rel} = 350 \Dim{km} \Dim{s^{-1}} \)
(upper line) and \( v \ind{rel} = 220 \Dim{km} \Dim{s^{-1}} \) (lower
line). Obviously, the conditions for the onset of ionization of the HVC's
boundaries caused by the CVE are satisfied for all data points shown in the
figure (see Eqs.~1 and 2). \\
Consequently, ionospheres similar to the ionospheric layers around planets and
comets form around the HVC's. 
Due to these ionospheres
more or less sharp boundaries form. Since the magnetic field is frozen in
the halo plasma, it will be draped around the HVC's during their motion through
the halo plasma. The convecting magnetic field generates currents in the
conducting ionospheres that keep the field from penetrating through the clouds,
at least initially. Depending on the conductivity and possibly local changes
of the halo field direction some parts of the field may eventually diffuse
into the clouds. In any case a more or less perfect magnetic cavity is created. The
draped field lines form a magnetic barrier. The magnetic field strength in such
a magnetic barrier may well exceed the value of \( 10 \mu \Dim{G} \) as can be
deduced from ram pressure equilibrium models (Kahn 1997). Closer to the centers
of the HVC's ionopauses form where the internal ionospheric pressures balance
the incident pressure of the halo. Beyond the magnetic barrier even a
collisionless bow shock may develop if the velocities of the HVC's
exceed the local halo magnetosonic velocity. The described scenario
for the formation of boundary layers around HVC's is illustrated in Fig.2.
It should be noted that the whole scenario resembles very much the well
studied interaction of the solar wind with unmagnetized planets (Luhmann 1986;
Luhmann et al. 1990) and comets (Formisano et al. 1982;
McComas et al. 1987; Flammer et al. 1997).

The Townsend condition for the production of secondary electrons that in course
are heated and ionize the neutral gas further defines a maximum radius for the
ionization region, the so-called CVE-radius:
\be
R \ind{CVE} \approx 1/n^{HVC} \sigma \ind{ion}  \approx 10^{-3} \Dim{pc}
\ee
where $\sigma \ind{ion} \approx 10^{-15} \Dim{cm^2}$ is the ionization cross section
(compare Flammer et al. 1997). \\
The given CVE-radius is very small compared to the the typical size of the
cloudlets in the MS which is of the order of \( 1 \Dim{kpc}\)
(Weiner \& Williams 1996). This means that newly created ions at the cloud
temperature should mainly be found in a boundary layer around the HVC. This
however is consistent with observations of localized H\(  \alpha \)-emission at
the leading edges of the clouds (Weiner \& Williams 1996). \\
Furthermore, the cloudlets should be stable against full ionization by a
Townsend discharge. \\
Diffuse H\( \alpha \)-emission from the inner parts of HVC's is then due to
remnants of the ionization front at earlier times.

It is worth mentioning that the
CVE also results in a braking of the
cold HVC's. The ionized boundaries interact with the magnetic field of
the halo that is mainly directed parallel to the Galactic plane.
The clouds are slowed down while the newly formed ions are accelerated.
The overall effect consists in a slowing down
of the ionization front.
Ultraviolet absorption observations hint at a deceleration of the HVC's
as they approach the Galactic disk (Benjamin 1999; Danly 1989). \\

\section{Discussion}

In our paper we discussed the critical velocity effect as a possible
explanation for the observed H$\alpha$ emission at the leading edges of cold
cloudlets in the MS. For suitable parameter regimes this
effect can yield  a rapid ionization of neutral gas streaming through
a magnetized plasma.

Our work has been motivated by intensive studies of the CVE
with respect to observed ionization phenomena in laboratory and space
plasmas. In laboratory plasmas ionization caused by the CVE
has been clearly identified in numerous experiments (e.g. Fahleson
1961; Danielsson \& Brenning 1975; Piel et al. 1980). The conditions, however,
under which the CVE becomes sufficiently efficient in
space plasmas are still under debate.

Ionization due to the CVE has been clearly identified in
context with a space experiment where an artificial cloud of barium was
released in the Earth's magnetosphere (Haerendel 1982). Besides, ion
production in the Io plasma torus around Jupiter and ionization in Io's
ionosphere are interpreted as the consequences of the CVE
(Cloutier et al. 1978; Galeev \& Chabibrachmanov 1983).

Further ionization phenomena in space which are discussed to be the
consequence of the CVE are the space shuttle glow
(Papadopoulos 1984) and, especially, ionization in cometary environments due
to the interaction between neutral gas sublimented from the cometary nuclei
and the magnetized solar wind plasma (Formisano et al. 1982).

All these examples yielding evidence for the importance of the
CVE refer to plasma - neutral gas interactions in laboratory
and space plasmas which differ from a large scale astrophysical system like
the MS in several aspects. However, the presented first
quantitative studies of the CVE considering parameter
regimes being typical for cloudlets in the MS strongly indicate
that the CVE can yield an efficient ionization of the
neutral gas at the boundaries of cold cloudlets in the halo plasma.

Thus, our kinematical studies have shown that the
CVE has to be considered as a possible explanation
for the observed H$\alpha$ emissions.
To get deeper insight into the question of to what extent the CVE
can contribute to the ionization of halo clouds detailed
dynamical multifluid studies are required.

\ \\
\ \\

{\sl Acknowledgments.}
This work has been supported by
the Deutsche Forschungsgemeinschaft through the grant LE 1039/5-1.

{}

\clearpage

\begin{figure}
\caption{The CVE in dependence of the magnetic field strength and the
particle density (cf. Eq.(\ref{3}) in double-logarithmic representation.
The CVE can operate efficiently
($\epsilon \approx 1$) for the parameter regime between the solid line
and the dotted one (for the case $v_{rel}=10^{-2}c$) or the solid line
and the dashed one (for the case $v_{rel}=10^{-4}c$), respectively.
The data points represent the parameters for HVC's in the MS.
\label{fig1}}
\end{figure}

\begin{figure}
\caption{ The formation of boundary layers around HVC's. The CVE leads
to the formation of an ionospheric boundary. Probably a collisionless
bow shock forms as in the case of magnetized and unmagnetized planets.
\label{fig2}}
\end{figure}

\end{document}